\documentclass[a4paper,twoside,twocolumn,english,amssymb,aps,prb,reprint,twocolumn,superscriptaddress,floats,showkeys,showpacs]{revtex4-1}

\usepackage[T1]{fontenc}
\usepackage[latin9]{inputenc}
\usepackage{babel}

\usepackage{amsmath}
\usepackage{amssymb}

\usepackage{graphicx}
\usepackage{color}
\usepackage[unicode=true,bookmarks=true,bookmarksnumbered=false,bookmarksopen=true,bookmarksopenlevel=1,breaklinks=true,pdfborder={0 0 0},backref=false,colorlinks=true,citecolor=blue]{hyperref}
\hypersetup{pdftitle={The Optical Activity of the Dark Exciton},pdfauthor={Yaroslav Don, Michal Zielinski and David Gershoni}}

\makeatletter

\pdfpageheight\paperheight
\pdfpagewidth\paperwidth 

\usepackage{braket}

\allowdisplaybreaks

\makeatother

\begin{document}

\title{The Optical Activity of the Dark Exciton} 

\author{Y. Don} 

\affiliation{The Physics Department and the Solid State Institute, Technion--Israel Institute of Technology, 32000 Haifa, Israel} 

\author{M. Zieli\'{n}ski} 

\affiliation{Institute of Physics, Faculty of Physics, Astronomy and Informatics, Nicolaus Copernicus University, ul.~Grudzi\k{a}dzka 5, PL-87-100 Toru\'{n}, Poland} 

\author{D. Gershoni}

\email{dg@physics.technion.ac.il}

\affiliation{The Physics Department and the Solid State Institute, Technion--Israel Institute of Technology, 32000 Haifa, Israel} 

\date{Jan. 2016}

%%%%%%%%%%%%%%%%%%%%%%%%%%%%%% Abstract

\begin{abstract}
	We present a phenomenological model to consider the effect of shape symmetry breaking on the optical properties of self-assembled quantum dots.  We compare between quantum dots with two-fold rotational and two reflections ($C_{2v}$) symmetry and quantum dots in which this symmetry is reduced by perturbation to one reflection only ($C_{s}$).  We show that this symmetry reduction drastically affects the optical activity of the dark exciton. 
	In symmetric quantum dots, one of the dark exciton eigenstate is totally dark and the other, due to heavy- and light-hole mixing, has a small dipole moment polarized along the symmetry axis (growth direction) of the quantum dot.  In non-symmetric quantum dots, the two dark excitons' eigenstates are mixed with the bright excitons' eigenstates which have cross-linearly polarized perpendicular to the growth direction dipole moments. As a result of this mixing one of the dark exciton eigenstate is dark while the other one does have dipole moment which is linearly polarized normal to the growth direction, like the lower energy bright exciton eigenstate. 
	Our model agrees well with recently obtained experimental data. 
\end{abstract}

\keywords{single quantum dot; micro-luminescence;} 

\pacs{78.67.Hc, 73.21.La, 78.55.Cr}

\maketitle

%%%%%%%%%%%%%%%%%%%%%%%%%%%%%% Bulk text 

\section{Introduction}

Excitons in single semiconductor quantum dots (QDs) play a central role in many schemes for applications in quantum optics and future quantum technologies~\cite{Loss.DiVincenzo.PRA.1998}.
QD confined excitons are generated by promoting one electron from the QD full valence band to the QD empty conduction band. If the electron spin is not altered in the process the generated excitons is optically active and it is called a bright exciton (BE). If however, the promoted electron spin is flipped (for example, during relaxation following non-resonant excitation) a dark exciton (DE), which is predominantly optically inactive, is formed. Since BEs are the fundamental optical excitations of these nanostructures they have been thoroughly studied both experimentally and theoretically. Their optical and coherent properties are quite well understood. Studies of DEs, which are as abundant as BEs in non-resonantly excited QDs, however, are relatively scarce. Recently, it was demonstrated that QD-confined DEs, despite their very weak optical activity, can still be efficiently accessed optically and electrically.~\cite{Poem.Nature.2010,Schwartz.PRX.2015,Schwartz.PRB.2015} Moreover, it was demonstrated that DEs form long lived two-level spin systems~\cite{Poem.Nature.2010} with very long coherence times~\cite{Schwartz.PRX.2015}. These naturally neutral, non-degenerate two level systems form matter qubits with obvious advantages~\cite{Schwartz.PRX.2015} over the single carrier spin qubits. It is therefore, important to study the DEs properties more thoroughly and to develop means for better understanding and thereby better controlling their properties. 

In a recent publication~\cite{Zielinski.PRB.2015} we used an atomistic model for studying the effect of QDs shape symmetry reduction on the optical properties of confined DEs in these QDs. We showed in Ref.~\cite{Zielinski.PRB.2015} that the deviation from symmetry, effectively mixes the DE states with the states of the BE. The mixing results in increased optical activity of the DEs, in agreement with recent experimental observations. 
Atomistic models, however, are very detailed, they consume time and large computing resources and usually produce results which are far from being intuitively understood.
 
Here we develop a simple, phenomenological model for studying the effect of the deviation from symmetry on the DE properties.
We show that despite its simplicity, our model does capture the essence of the symmetry reduction induced BE-DE mixing. It thereby provides an intuitive analytical tool for quantitative studies of QD confined DEs. 

Theoretical studies of the fine structure of the confined exciton
in semiconductor quantum dots (QDs) epitaxially grown on $[001]$
oriented substrate generally assume combined lattice and structural
symmetry of $C_{2v}$ (i.~e., symmetry under rotations of $\pi$
radians around the structural symmetry axis $[001]$, and under two reflections about perpendicular planes 
which contain the symmetry axis: the $[110]\textrm{--}[001]$ and the $[1\bar{1}0]\textrm{--}[001]$ planes.~\cite{Ivchenko.Book.2005,Takagahara.PRB.2000,Goupalov.JETP.1998}.
Since the quantum size effect and the strain result in large energy
difference between the heavy-holes with total angular momentum projection
$\pm3/2$ on the symmetry axis and the light-holes with total angular
momentum projection of $\pm1/2$, the lowest energy exciton states
are composed mainly of four different angular momentum configurations
of electron hole pairs. Two exciton states, in which the electron
and the heavy-hole have parallel angular momentum projections $\ket{+2},\ket{-2}$
with vanishing dipole matrix to optical transitions, called dark excitons
(DEs); and two states, in which their spin projections are antiparallel,
$\ket{+1},\ket{-1}$, forming the fundamental optical excitations
of the QD, and therefore termed bright excitons (BEs).

General theoretical considerations, based on group theory arguments,
imply that excitons in semiconductor nanostructures of $C_{2v}$ symmetry have four lowest energy eigenstates. Two of which have cross-linearly
in-plane polarized dipole moments, associated with the two planes of reflection.  
One eigenstate have dipole moment polarized along the vertical symmetry axis ($\hat{z}$
direction, or $[001]$), and one eigenstate, which is completely dark. The first two eigenstates are associated with the bright excitons, while the later two are associated with the dark excitons.~\cite{Ivchenko.Pikus.Book.1997,Dupertuis.PRL.2011}
More specific considerations, which take into account the nature of the valence band structure in semiconductors, show that the $[001]$ polarized optical activity of one of the DE eigenstates and the lack of activity of the other eigenstate are attributed to constructive and destructive contributions to their dipole moments, respectively, due to heavy-light hole mixing.~\cite{Takagahara.PRB.2000,Lovett.PRB.2005}

Below, motivated by recent experimental observations, we construct a simple model 
which quantitatively account for the changes in the optical activity of the DE
induced by small deviations from the exact  $C_{2v}$ symmetry.    
 
\section{Theoretical Model and Calculations}

The electron hole exchange part of the Coloumb interaction removes
the degeneracy between the 4 lowest exciton states. From general symmetry
considerations, it can be shown that for $C_{2v}$ QDs the exchange
interaction Hamiltonian written in the base $\ket{+1},\ket{-1},\ket{+2},\ket{-2}$
has the following form~\cite{Ivchenko.Book.2005,Takagahara.PRB.2000,Goupalov.JETP.1998}:
\begin{equation}
	\mathcal{H}_{C_{2v}}=
	\frac{1}{2}\begin{pmatrix}
		\Delta_{0} & \Delta_{1}^{*} & 0 & 0\\
		\Delta_{1} & \Delta_{0} & 0 & 0\\
		0 & 0 & -\Delta_{0} & \Delta_{2}^{*}\\
		0 & 0 & \Delta_{2} & -\Delta_{0}
	\end{pmatrix}
	\label{eq:Hexch-simple}
\end{equation}
Here $\Delta_{i},\,i=0,1,2$ are parameters that one either measures~\cite{AlonBraitbart.PhysE.2006,Bayer.PRB.2002,Bayer.PRB.2000,Poem.Nature.2010,Schwartz.PRX.2015}
or try to calculate using simplified models~\cite{Ivchenko.Pikus.Book.1997,Ivchenko.Book.2005,Goupalov.JETP.1998,Poem.PRB.2007}.

For a $C_{2v}$ symmetrical QD there is no mixing between the DEs
and the BEs eigenstates. The two subspaces are energetically separated
by $\Delta_{0}$. The parameter $\Delta_{1}$, which removes the degeneracy
between the two BE states, is closely related to the oscillator strength
for optical transitions to these fundamental excitations~\cite{Poem.PRB.2007,Takagahara.PRB.2000}.
For instance, using the envelope function approximation $\Delta_{1}$
is given by~\cite{Takagahara.PRB.2000} 
\begin{equation}
	\begin{aligned}
		\tfrac{1}{2}\Delta_{1}=\Bra{\Phi_{\mathrm{h}}^{\Uparrow}\left(\mathbf{r}_{1}\right)\Phi_{\mathrm{e}}^{\downarrow}\left(\mathbf{r}_{2}\right)}\tfrac{e^{2}}{\epsilon}\tfrac{\boldsymbol{\mu}_{\downarrow\Uparrow}^{\dagger}\left(I-3\hat{n}\otimes\hat{n}^{\dagger}\right)\boldsymbol{\mu}_{\uparrow\Downarrow}}{\left|\mathbf{r}_{1}-\mathbf{r}_{2}\right|^{3}}\\
		\Ket{\Phi_{\mathrm{e}}^{\uparrow}\left(\mathbf{r}_{1}\right)\Phi_{\mathrm{h}}^{\Downarrow}\left(\mathbf{r}_{2}\right)}
	\end{aligned}
	\label{eq:long-range-exch}
\end{equation}
where $\epsilon$ is the dielectric constant, $e$ is the electronic charge,
$\Phi_{\mathrm{e(h)}}^{s_{\mathrm{e}}(s_{\mathrm{h}})}$
is the electron (heavy-hole) conduction-band (valence-band) envelope
function with spin $s_{\mathrm{e}}\left(s_{\mathrm{h}}\right)$,
$\mathbf{r}_{1}$, $\mathbf{r}_{2}$ are the two carriers position vectors,
$\hat{n}$ is a unit vector in the direction of $\mathbf{r}_{1}-\mathbf{r}_{2}$,
$I$ is the $3\times3$ unit matrix, $\hat{n}\otimes\hat{n}^{\dagger}$
a dyadic product, 
and
$\boldsymbol{\mu}_{s_{\mathrm{e}}s_{\mathrm{h}}}$ is the valence-conduction
band dipole matrix element. 

The relation between the dipole matrix
element and the momentum matrix element is given by~\cite{Poem.PRB.2007,Takagahara.PRB.2000}
\begin{equation}
	\boldsymbol{\mu}_{s_{e}s_{h}}=\frac{-\mathrm{i}\hbar}{m_{0}E_{g}}\mathbf{M}_{s_{e}s_{h}}
\end{equation}
where $E_{g}$ is the bandgap energy. The momentum matrix elements
are~\cite{Ivchenko.Book.2005,Takagahara.PRB.2000} 
\begin{subequations}
	\label{eq:BE-DE-matrix-elements} 
	\begin{align}
		\mathbf{M}_{\downarrow\Uparrow} & =\tfrac{\mathrm{i}}{2}\sqrt{m_{0}E_{P}}\left(1,-\mathrm{i},0\right)\label{eq:BEp1-matrix-elements}\\
		\mathbf{M}_{\uparrow\Downarrow} & =\tfrac{\mathrm{i}}{2}\sqrt{m_{0}E_{P}}\left(1,\mathrm{i},0\right)\label{eq:BEm1-matrix-elements}\\
		\mathbf{M}_{\uparrow\Uparrow}=\mathbf{M}_{\downarrow\Downarrow} & =0\label{eq:DE-matrix-elements}
	\end{align}
\end{subequations}
where $E_{P}$ is Kane's energy.~\cite{Takagahara.PRB.2000} 

$\Delta_{1,2}$ are in general complex numbers~\cite{Ivchenko.Book.2005}
and can be expressed as $\Delta_{1,2}=\delta_{1,2}\mathrm{e}^{2\mathrm{i}\theta_{1,2}}$,
where $\delta_{1,2}$ are positive numbers. Thus the eigenvalues of
the Hamiltonian are expressed as 
\begin{subequations} 
	\begin{align}
		E_{\mathrm{BE}_{\pm}} & =\tfrac{1}{2}(\Delta_{0}\pm\delta_{1})\\
		E_{\mathrm{DE}_{\pm}} & =\tfrac{1}{2}(-\Delta_{0}\pm\delta_{2})
	\end{align}
	\label{eq:symmetricEV} 
\end{subequations}
and the eigenvectors as 
\begin{subequations} 
	\begin{align}
		v_{\mathrm{BE}_{\pm}} & = 
		\frac{1}{\sqrt{2}}\begin{pmatrix}
			\mathrm{e^{-i\theta_{1}}}\\
			\pm\mathrm{e^{i\theta_{1}}}
		\end{pmatrix} \\
		v_{\mathrm{DE}_{\pm}} & =
		\frac{1}{\sqrt{2}}\begin{pmatrix}
			\mathrm{e^{-i\theta_{2}}}\\
			\pm\mathrm{e^{i\theta_{2}}}
		\end{pmatrix}
	\end{align}
	\label{eq:symmetricEF} 
\end{subequations}

Using the expressions for the momentum matrix element in Eqs.~\eqref{eq:BEp1-matrix-elements}
and \eqref{eq:BEm1-matrix-elements} one finds that the positive ($v_{\mathrm{BE_{+}}}$)
and negative ($v_{\mathrm{BE_{-}}}$) eigenstates of the BE have dipole
matrix elements linearly polarized along the $(\cos\theta_{1},-\sin\theta_{1},0)$
and $(\sin\theta_{1},\cos\theta_{1},0)$ directions, respectively,
where $\theta_{1}$ is measured from the $[100]$ crystallographic
direction.

Atomistic calculations~\cite{Korkusinski.PRB.2013,Zielinski.PRB.2015}
and accumulated experimental data~\cite{Young.Shields.PRB.2005,Langbein.PRB.2004,Bimberg.Book.1999,Schwartz.PRX.2015}
imply that most often the lowest (highest) energy BE emission spectral
line is polarized along the $[1\bar{1}0]$ ($[110]$) direction, even
for a circularly symmetric QD. If one defines the lowest energy line
polarization as horizontal polarization (i.e symmetrical superposition
of right and left hand circular polarizations) this situation is described
by $\theta_{1}=135^{\circ}$.

The value of $\Delta_{2}$ is mostly determined by the short range
e-h exchange interaction~\cite{Ivchenko.Pikus.Book.1997} which has
the symmetry of the unit cell. This implies that $\Delta_{2}$ must
be a real number~\cite{Bayer.PRB.2002}, thus compelling $\theta_{2}$
to be either $0^{\circ}$ or $90^{\circ}$. Atomistic model simulations~\cite{Zielinski.PRB.2015,Bryant.PRL.2010},
as well as recent experimental data~\cite{Schwartz.PRX.2015} indicate
that $\theta_{2}=90^{\circ}$.

From Eq.~\eqref{eq:DE-matrix-elements} it follows that the DEs are
completely dark. However, if one allows some residual heavy-hole light-hole
mixing it follows that one of the DE eigenstates has small $\hat{z}$-polarized
dipole moment, while the other one is totally dark~\cite{Ivchenko.Pikus.Book.1997,Ivchenko.Book.2005,Dupertuis.PRL.2011}.
Realistic atomistic model calculations of InAs/GaAs self assembled
QDs indeed result with 3--6 orders of magnitude weaker $\hat{z}$-polarized
optical activity of one of the DE eigenstate and a much weaker activity
of the other DE eigenstate~\cite{Korkusinski.PRB.2013,Zielinski.JPCM.2013,Smolenski.PRB.2012,Zielinski.PRB.2015}.

In reality, ideally symmetrized systems of macroscopic scale are extremely
rare. Recent theoretical studies of epitaxial growth of strained heterostructures~\cite{Spencer.PRB.2013}
show that indeed self-assembled QDs can actually grow highly asymmetrical,
largely deviating from $C_{2v}$ symmetry. In an asymmetrical QD,
the subspaces of the BEs and DEs are no longer separated and their
eigenstates are mixed~\cite{Bayer.PRB.2002}.

In order to methodically study the effects of the structural symmetry
reduction of the QD on the excitons, in our recent work~\cite{Zielinski.PRB.2015} 
we used atomistic model, in which an inclined planar facet was introduced
between the QD and the covering host material, thereby reducing
the symmetry of the QD. As a result of the symmetry reduction both
the electron and the hole have non-vanishing in-plane spin projection
expectation values, where for a $C_{2v}$ symmetrical QD these expectation
values vanish.

Here, we model the symmetry reduction by introducing a small angle
$\varphi$ by which the symmetry axis of the QD is tilted relative
to the $[001]$ crystallographic direction. 
As a result, the quantization axis of the QD potential is no longer aligned with the underlying semiconductor lattice, which defines
the momentum matrix elements in Eq.~\eqref{eq:BE-DE-matrix-elements}.
The electron (heavy-hole) envelope wavefunction's symmetry axis is therefore inclined by an angle $\varphi_{\mathrm{e}}$
($\varphi_{\mathrm{h}}$) relative to the $[001]$ crystallographic direction. 
Effectively, the new projections of the carrier spins on the envelope wavefunctions symmetry axes are given by: 
\begin{subequations} 
	\begin{align}
		\begin{pmatrix}\tilde{\uparrow}\\
			\tilde{\downarrow}
		\end{pmatrix} & =\begin{pmatrix}\cos\varphi_{\mathrm{e}} & \sin\varphi_{\mathrm{e}}\\
			-\sin\varphi_{\mathrm{e}} & \cos\varphi_{\mathrm{e}}
		\end{pmatrix}\begin{pmatrix}\uparrow\\
			\downarrow
		\end{pmatrix}\\
		\begin{pmatrix}\tilde{\Uparrow}\\
			\tilde{\Downarrow}
		\end{pmatrix} & =\begin{pmatrix}\cos\varphi_{\mathrm{h}} & \sin\varphi_{\mathrm{h}}\\
			-\sin\varphi_{\mathrm{h}} & \cos\varphi_{\mathrm{h}}
		\end{pmatrix}\begin{pmatrix}\Uparrow\\
			\Downarrow
		\end{pmatrix}
	\end{align}
\end{subequations}
As a result the matrix elements of Eqs.~\eqref{eq:BE-DE-matrix-elements}
transform to 
\begin{subequations} 
	\begin{align}
		\mathbf{M}_{\widetilde{\downarrow\Uparrow}} & =+\mathbf{M}_{\downarrow\Uparrow}\cos\varphi_{\mathrm{e}}\cos\varphi_{\mathrm{h}}-\mathbf{M}_{\uparrow\Downarrow}\sin\varphi_{\mathrm{e}}\sin\varphi_{\mathrm{h}}\\
		\mathbf{M}_{\widetilde{\uparrow\Downarrow}} & =-\mathbf{M}_{\downarrow\Uparrow}\sin\varphi_{\mathrm{e}}\sin\varphi_{\mathrm{h}}+\mathbf{M}_{\uparrow\Downarrow}\cos\varphi_{\mathrm{e}}\cos\varphi_{\mathrm{h}}\\
		\mathbf{M}_{\widetilde{\uparrow\Uparrow}} & =+\mathbf{M}_{\downarrow\Uparrow}\sin\varphi_{\mathrm{e}}\cos\varphi_{\mathrm{h}}+\mathbf{M}_{\uparrow\Downarrow}\cos\varphi_{\mathrm{e}}\sin\varphi_{\mathrm{h}}\\
		\mathbf{M}_{\widetilde{\downarrow\Downarrow}} & =-\mathbf{M}_{\downarrow\Uparrow}\cos\varphi_{\mathrm{e}}\sin\varphi_{\mathrm{h}}-\mathbf{M}_{\uparrow\Downarrow}\sin\varphi_{\mathrm{e}}\cos\varphi_{\mathrm{h}}
	\end{align}
\end{subequations}
Using only first order terms in $\varphi_e$ and $\varphi_h$ these equations become
\begin{subequations} 
	\begin{align}
		\mathbf{M}_{\widetilde{\downarrow\Uparrow}} & =\mathbf{M}_{\downarrow\Uparrow}\\
		\mathbf{M}_{\widetilde{\uparrow\Downarrow}} & =\mathbf{M}_{\uparrow\Downarrow}\\
		\mathbf{M}_{\widetilde{\uparrow\Uparrow}} & =\mathbf{M}_{\downarrow\Uparrow}\cdot\varphi_{\mathrm{e}}+\mathbf{M}_{\uparrow\Downarrow}\cdot\varphi_{\mathrm{h}}\label{eq:DEp2-rot-matrix-element}\\
		\mathbf{M}_{\widetilde{\downarrow\Downarrow}} & =-\left(\mathbf{M}_{\downarrow\Uparrow}\cdot\varphi_{\mathrm{h}}+\mathbf{M}_{\uparrow\Downarrow}\cdot\varphi_{\mathrm{e}}\right)\label{eq:DEm2-rot-matrix-element}
	\end{align}
\end{subequations}
By adding and subtracting Eq.~\eqref{eq:DEm2-rot-matrix-element}
to and from Eq.~\eqref{eq:DEp2-rot-matrix-element} one gets: 
\begin{subequations} 
	\begin{align}
		\mathbf{M}_{\widetilde{\uparrow\Uparrow}}+\mathbf{M}_{\widetilde{\downarrow\Downarrow}} & =(\mathbf{M}_{\downarrow\Uparrow}-\mathbf{M}_{\uparrow\Downarrow})\cdot(\varphi_{\mathrm{e}}-\varphi_{\mathrm{h}})\\
		\mathbf{M}_{\widetilde{\uparrow\Uparrow}}-\mathbf{M}_{\widetilde{\downarrow\Downarrow}} & =(\mathbf{M}_{\downarrow\Uparrow}+\mathbf{M}_{\uparrow\Downarrow})\cdot(\varphi_{\mathrm{e}}+\varphi_{\mathrm{h}})
	\end{align}
	\label{eq:symmetric-matrix-elements} 
\end{subequations}

Eqs.~\eqref{eq:symmetric-matrix-elements} imply that the symmetric
(antisymmetric) DE eigenstate is coupled only to the anti-symmetric
(symmetric) BE eigenstate and that the coupling constant is proportional
to $\varphi_{\mathrm{e}}-\varphi_{\mathrm{h}}$ ($\varphi_{\mathrm{e}}+\varphi_{\mathrm{h}}$).

We proceed by using Eq.~\eqref{eq:long-range-exch}, which associates the momentum matrix elements with the long range 
exchange interaction mixing term $\Delta_1$, to obtain the non-diagonal mixing
terms between the BE and DE eigenstates of the Hamiltonian of a $C_{2v}$
QD as expressed by Eqs.~\eqref{eq:symmetricEV} and \eqref{eq:symmetricEF}. 
Substituting the modified momentum matrix elements of Eqs.~\eqref{eq:symmetric-matrix-elements} into Eq.~\eqref{eq:long-range-exch} one gets the following modified Hamiltonian:

\begin{widetext}
	\begin{equation}
		\mathcal{H}'_{C_{s}}=
		\frac{1}{2}\begin{pmatrix}
			\Delta_{0}+\delta_{1} & 0 & \delta_{1}\cdot\left(\varphi_{\mathrm{h}}-\varphi_{\mathrm{e}}\right) & 0\\
			0 & \Delta_{0}-\delta_{1} & 0 & \delta_{1}\cdot\left(\varphi_{\mathrm{h}}+\varphi_{\mathrm{e}}\right)\\
			\delta_{1}\cdot\left(\varphi_{\mathrm{h}}-\varphi_{\mathrm{e}}\right) & 0 & -\Delta_{0}+\delta_{2} & 0\\
			0 & \delta_{1}\cdot\left(\varphi_{\mathrm{h}}+\varphi_{\mathrm{e}}\right) & 0 & -\Delta_{0}-\delta_{2}
		\end{pmatrix}
		\label{eq:Hexch-full-diag}
	\end{equation}
	This Hamiltonian can be expressed in terms of the original
	base $\ket{+1},\ket{-1},\ket{+2},\ket{-2}$: 
	\begin{equation}
		\mathcal{H}_{C_{s}}=
		\frac{1}{2}\begin{pmatrix}
			\Delta_{0} & \delta_{1}\mathrm{e^{-2i\theta_{1}}} & \mathrm{i}\delta_{1}\mathrm{e^{-i\theta_{1}+i\theta_{2}}}\cdot\varphi_{\mathrm{h}} & -\mathrm{i}\delta_{1}\mathrm{e^{-i\theta_{1}-i\theta_{2}}}\cdot\varphi_{\mathrm{e}}\\
			\delta_{1}\mathrm{e^{2i\theta_{1}}} & \Delta_{0} & -\mathrm{i}\delta_{1}\mathrm{e^{i\theta_{1}+i\theta_{2}}}\cdot\varphi_{\mathrm{e}} & \mathrm{i}\delta_{1}\mathrm{e^{i\theta_{1}-i\theta_{2}}}\cdot\varphi_{\mathrm{h}}\\
			-\mathrm{i}\delta_{1}\mathrm{e^{i\theta_{1}-i\theta_{2}}}\cdot\varphi_{\mathrm{h}} & \mathrm{i}\delta_{1}\mathrm{e^{-i\theta_{1}-i\theta_{2}}}\cdot\varphi_{\mathrm{e}} & -\Delta_{0} & \delta_{2}\mathrm{e^{-2i\theta_{2}}}\\
			\mathrm{i}\delta_{1}\mathrm{e^{i\theta_{1}+i\theta_{2}}}\cdot\varphi_{\mathrm{e}} & -\mathrm{i}\delta_{1}\mathrm{e^{-i\theta_{1}+i\theta_{2}}}\cdot\varphi_{\mathrm{h}} & \delta_{2}\mathrm{e^{2i\theta_{2}}} & -\Delta_{0}
		\end{pmatrix}\label{eq:Hexch-full}
	\end{equation} 

	The new eigenenergies and eigenvectors of the reduced symmetry Hamiltonian
	of Eq.~\eqref{eq:Hexch-full-diag} are now: 
	\begin{subequations} 
		\begin{align}
			E_{1}=\tfrac{+\delta_{1}-\delta_{2}}{4}+\tfrac{1}{2}\sqrt{\bigl(\tfrac{2\Delta_{0}+\delta_{1}+\delta_{2}}{2}\bigr)^{2}+\delta_{1}^{2}\bigl(\varphi_{\mathrm{h}}-\varphi_{\mathrm{e}}\bigr)^{2}}, & \quad v_{1}=N_{1}\Bigl(\tfrac{2\Delta_{0}+\delta_{1}+\delta_{2}}{2\delta_{1}\left(\varphi_{\mathrm{h}}-\varphi_{\mathrm{e}}\right)}+\sqrt{\bigl(\tfrac{2\Delta_{0}+\delta_{1}+\delta_{2}}{2\delta_{1}\left(\varphi_{\mathrm{h}}-\varphi_{\mathrm{e}}\right)}\bigr)^{2}+1},0,1,0\Bigr)\\
			E_{2}=\tfrac{-\delta_{1}+\delta_{2}}{4}+\tfrac{1}{2}\sqrt{\bigl(\tfrac{2\Delta_{0}-\delta_{1}-\delta_{2}}{2}\bigr)^{2}+\delta_{1}^{2}\bigl(\varphi_{\mathrm{h}}+\varphi_{\mathrm{e}}\bigr)^{2}}, & \quad v_{2}=N_{2}\Bigl(0,\tfrac{2\Delta_{0}-\delta_{1}-\delta_{2}}{2\delta_{1}\left(\varphi_{\mathrm{h}}+\varphi_{\mathrm{e}}\right)}+\sqrt{\bigl(\tfrac{2\Delta_{0}-\delta_{1}-\delta_{2}}{2\delta_{1}\left(\varphi_{\mathrm{h}}+\varphi_{\mathrm{e}}\right)}\bigr)^{2}+1},0,1\Bigr)\\
			E_{3}=\tfrac{-\delta_{1}+\delta_{2}}{4}-\tfrac{1}{2}\sqrt{\bigl(\tfrac{2\Delta_{0}-\delta_{1}-\delta_{2}}{2}\bigr)^{2}+\delta_{1}^{2}\bigl(\varphi_{\mathrm{h}}-\varphi_{\mathrm{e}}\bigr)^{2}}, & \quad v_{3}=N_{3}\Bigl(\tfrac{2\Delta_{0}-\delta_{1}-\delta_{2}}{2\delta_{1}\left(\varphi_{\mathrm{h}}-\varphi_{\mathrm{e}}\right)}-\sqrt{\bigl(\tfrac{2\Delta_{0}-\delta_{1}-\delta_{2}}{2\delta_{1}\left(\varphi_{\mathrm{h}}-\varphi_{\mathrm{e}}\right)}\bigr)^{2}+1},0,1,0\Bigr)\\
			E_{4}=\tfrac{+\delta_{1}-\delta_{2}}{4}-\tfrac{1}{2}\sqrt{\bigl(\tfrac{2\Delta_{0}+\delta_{1}+\delta_{2}}{2}\bigr)^{2}+\delta_{1}^{2}\bigl(\varphi_{\mathrm{h}}+\varphi_{\mathrm{e}}\bigr)^{2}}, & \quad v_{4}=N_{4}\Bigl(0,\tfrac{2\Delta_{0}+\delta_{1}+\delta_{2}}{2\delta_{1}\left(\varphi_{\mathrm{h}}+\varphi_{\mathrm{e}}\right)}-\sqrt{\bigl(\tfrac{2\Delta_{0}+\delta_{1}+\delta_{2}}{2\delta_{1}\left(\varphi_{\mathrm{h}}+\varphi_{\mathrm{e}}\right)}\bigr)^{2}+1},0,1\Bigr)
		\end{align}
	\end{subequations}
	where $N_{i}$ are normalization constants. Noting that $\Delta_{0}\gg\delta_{1}\gg\delta_{2}$
	and $|\varphi_{\mathrm{e}}|, |\varphi_{\mathrm{h}}| \ll 1$, we may approximate
	those equations as: 
	\begin{subequations}
		\label{eqs:BE-DE-eigenstates} 
		\begin{align}
			E_{1}=\frac{+\Delta_{0}+\delta_{1}}{2}+\frac{\delta_{1}^{2}\bigl(\varphi_{\mathrm{h}}-\varphi_{\mathrm{e}}\bigr)^{2}}{4\Delta_{0}}, & \qquad v_{1}=N_{-}\Bigl(1,0,+\frac{\delta_{1}\cdot\bigl(\varphi_{\mathrm{h}}-\varphi_{\mathrm{e}}\bigr)}{2\Delta_{0}},0\Bigr)\\
			E_{2}=\frac{+\Delta_{0}-\delta_{1}}{2}+\frac{\delta_{1}^{2}\bigl(\varphi_{\mathrm{h}}+\varphi_{\mathrm{e}}\bigr)^{2}}{4\Delta_{0}}, & \qquad v_{2}=N_{+}\Bigl(0,1,0,+\frac{\delta_{1}\cdot\bigl(\varphi_{\mathrm{h}}+\varphi_{\mathrm{e}}\bigr)}{2\Delta_{0}}\Bigr)\\
			E_{3}=\frac{-\Delta_{0}+\delta_{2}}{2}-\frac{\delta_{1}^{2}\bigl(\varphi_{\mathrm{h}}-\varphi_{\mathrm{e}}\bigr)^{2}}{4\Delta_{0}}, & \qquad v_{3}=N_{-}\Bigl(-\frac{\delta_{1}\cdot\bigl(\varphi_{\mathrm{h}}-\varphi_{\mathrm{e}}\bigr)}{2\Delta_{0}},0,1,0\Bigr)\\
			E_{4}=\frac{-\Delta_{0}-\delta_{2}}{2}-\frac{\delta_{1}^{2}\bigl(\varphi_{\mathrm{h}}+\varphi_{\mathrm{e}}\bigr)^{2}}{4\Delta_{0}}, & \qquad v_{4}=N_{+}\Bigl(0,-\frac{\delta_{1}\cdot\bigl(\varphi_{\mathrm{h}}+\varphi_{\mathrm{e}}\bigr)}{2\Delta_{0}},0,1\Bigr)
		\end{align}
	\end{subequations} 
	where $N_{\pm}$ are normalization constants. 
\end{widetext}

\begin{figure}
	\centering{}
	\includegraphics[width=0.95\columnwidth]{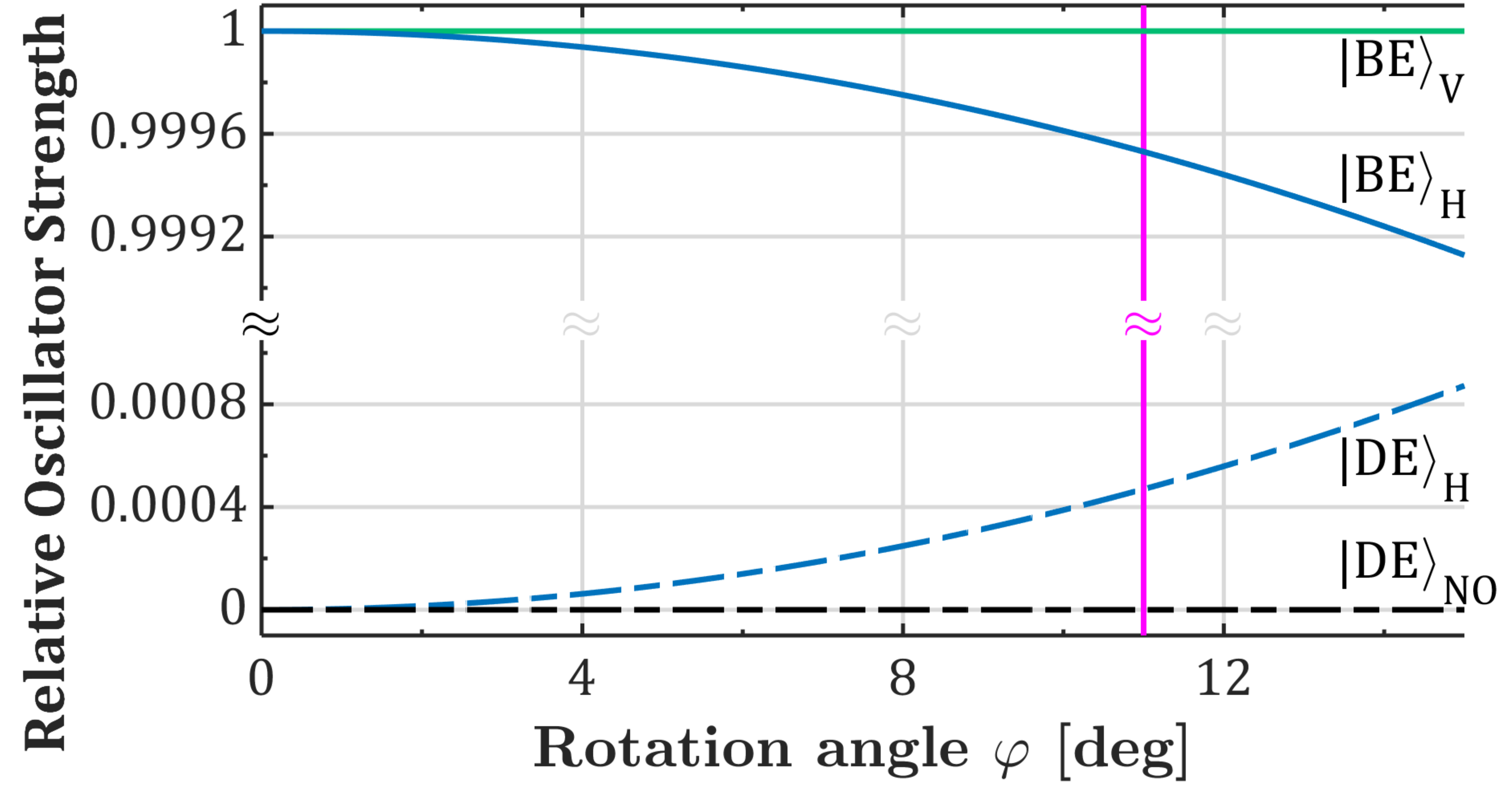}
	\caption{The calculated relative magnitude of the oscillator strength of the
		4 exciton eigenstates (Eq.~\eqref{eq:Hexch-full-diag} for $\varphi_{\mathrm{h}}=\varphi_{\mathrm{e}}\equiv\varphi$).
		Solid $\mathrm{\protect\ket{BE}_{V/H}}$ (dash $\mathrm{\protect\ket{DE}_{H/NO}}$)
		lines represent BE (DE) eigenstates. Blue, green, and black colors
		represent $H$, $V$ polarizations, and no optical activity, respectively.
		The experimentally measured values of Ref.~\cite{Schwartz.PRX.2015} are indicated by the magenta vertical line. 
	\label{fig:relative-OS} }
\end{figure}

To proceed, we now discuss the model angles $\varphi_{\mathrm{h}}$ and $\varphi_{\mathrm{e}}$,
which for a $C_{\mathrm{2v}}$ symmetric QD are $\varphi_{\mathrm{e}}=\varphi_{\mathrm{h}}=0$.
However, for a slightly lower symmetry QD ($C_{\mathrm{s}}$), 
the QD confined electron and hole are in their respective ground states,
their envelope wavefunctions are typically restricted to the QD volume, and both possess the same in plane ``$s$-like'' symmetry. 
It is therefore expected that the respective angles of inclination between the electron and the hole 
envelope wavefunction symmetry axes and the crystallographic $[001]$ 
direction are similar. Both are approximately equal to $\varphi$: $\varphi_{\mathrm{h}}\simeq\varphi_{\mathrm{e}}\simeq\varphi$.
By substituting the approximation $\varphi_{\mathrm{h}}=\varphi_{\mathrm{e}}\equiv\varphi$,
in Eq.~\eqref{eq:Hexch-full-diag} one immediately sees that only the lower energy DE eigenstate acquires
optical activity by mixing with the lower energy eigenstate of the
BE. The dipole moment of this weakly visible DE eigenstate is thus
polarized like that of the BE eigenstate, with which it is mixed ($H$-polarized), as indeed
was recently observed experimentally~\cite{Schwartz.PRX.2015}. For
a positive $\delta_{2}$ and $\theta_{2}=\frac{\pi}{2}$ this DE eigenstate
is antisymmetric under electron-hole exchange, also in agreement with the experimental observation.~\cite{Schwartz.PRX.2015} 
The energy and eigenvector
of the visible DE are given by: 
\begin{equation}
	\begin{aligned}
		E_{4} & \simeq\frac{-\Delta_{0}-\delta_{2}}{2}-\frac{\delta_{1}^{2}}{\Delta_{0}}\cdot\varphi^{2}\\
		v_{4} & \simeq\Bigl(0,\frac{-\delta_{1}\cdot\varphi}{\Delta_{0}},0,1\Bigr)
	\end{aligned}
\end{equation}
The calculated relative magnitude of the oscillator strength of the
4 exciton eigenstates as a function of the symmetry reduction $\varphi$ 
is presented in Fig.~\ref{fig:relative-OS}.
Our simple model can be readily compared with the recently measured
experimental data of Ref.~\cite{Schwartz.PRX.2015} where the visible
DE to BE oscillator strength ratio of $\sim5\times10^{-4}$ is obtained
at mixing strength of $\varphi=11^{\circ}$. The obtained angle agrees
with the facet inclination angle of the atomistic model of Ref.~\cite{Zielinski.PRB.2015}.
  
If, however, the exciton is formed between electron and hole of different levels, 
the inclination angles $\varphi_{\mathrm{h}}$ and $\varphi_{\mathrm{e}}$ are not expected to be the same. 
Intuitively, one expects that the envelope wavefunction of a higher energy carrier, will be less restricted to the QD volume and thus less affected by the QD deviation from symmetry. 
Thus for example, for a DE between ground state electron and excited state heavy-hole one expects $|\varphi_{\mathrm{h}}| \ll |\varphi_{\mathrm{e}}|$. In this case, both DE eigenstates will be optically active with almost equal in plane polarized dipole moments, as indeed was recently observed experimentally.~\cite{Schwartz.PRB.2015}

%%%%%%%%%%%%%%%%%%%%%%%%%%%%%% Summary

\section{Conclusions}

We use a phenomenological model to examine the effects of symmetry
reduction on the optical properties of the excitons in self
assembled semiconductor QDs. We compare between excitons in $C_{2v}$ symmetrical 
QDs and excitons in QDs with small deviations from this symmetry.  
We model the symmetry reduction by an angle of inclination 
between the quantum dot symmetry axis and the crystallographic directions. 
We show that while the reduction in symmetry barely affects 
the bright exciton eigenstates and their optical activity,
it strongly affects the optical activity of the dark exciton by mixing its eigenstates
with these of the bright exciton.  
For the ground state dark exciton the lowest energy eigenstate has
in-plane dipole moment which is polarized like the lowest energy bright exciton eigenstate, but the other eigenstate 
has much weaker optical activity. For excited dark exciton eigenstates this strong anisotropy in the in-plane dipole moment
strengths is greatly reduced. 
The polarization selection rules, the oscillator strengths ratio, and
the excitonic energy levels order, are well compared with recently measured data. 

\begin{acknowledgments}
	The support of the Israeli Science Foundation (ISF), the Technion's RBNI and the Israeli Focal Technology Area on ``Nanophotonics for Detection'' is gratefully acknowledged. 
	MG acknowledges support from the Polish Ministry of Science and Higher Education (research project No IP 2012064572, Iuventus Plus).
	We also thank Joseph Avron for useful discussions. 
\end{acknowledgments}

%%%%%%%%%%%%%%%%%%%%%%%%%%%%%% Backmatter

\bibliographystyle{aipnum4-1}
\bibliography{Optical_Activity_of_DE_-_Refs}

\end{document}